\documentclass[12pt]{article}

\pdfoutput=1

\usepackage[left=3cm,top=2cm,right=3cm,bottom=3cm]{geometry}

\usepackage{amsmath}
\usepackage{amssymb}
\usepackage{latexsym}
\usepackage{pstricks}
\usepackage{amsfonts}
\usepackage{esint}
\usepackage{algorithm}

\usepackage{graphicx}
\usepackage{epstopdf}

\usepackage{paralist}
\usepackage[colorlinks=true]{hyperref}

\hypersetup{urlcolor=blue, citecolor=red}

\newcommand{\ds}{\displaystyle}

\newcommand{\Frac}[2]{\ds \frac{#1}{#2}}
\newcommand{\eref}[1]{(\ref{#1})}

\title{Vectorized and Parallel Particle Filter SMC\\Parameter Estimation for Stiff ODEs}

\author{Andrea Arnold$^{1,2}$, Daniela Calvetti$^3$ and Erkki Somersalo$^3$}

\date{
{\footnotesize
$^1$ Center for Quantitative Sciences in Biomedicine\\
North Carolina State University, Campus Box 8213, 2700 Stinson Drive, 308 Cox Hall\\
Raleigh, NC 27695-8213, USA\\
\medskip
$^2$ Department of Mathematics\\
North Carolina State University, Campus Box 8205, 2311 Stinson Drive, 2108 SAS Hall\\
Raleigh, NC 27695-8205, USA\\
\medskip
$^3$ Department of Mathematics, Applied Mathematics and Statistics\\
Case Western Reserve University, 10900 Euclid Ave.\\
Cleveland, OH 44106, USA\\
\medskip
E-mail: anarnold@ncsu.edu, dxc57@case.edu, ejs49@case.edu
}
}

\begin{document}
\maketitle

% Abstract
\begin{abstract}
Particle filter (PF) sequential Monte Carlo (SMC) methods are very attractive for the estimation of parameters of time dependent systems where the data is either not all available at once, or the range of time constants is wide enough to create problems in the numerical time propagation of the states. The need to evolve a large number of particles makes PF-based methods computationally challenging, the main bottlenecks being the time propagation of each particle and the large number of particles. While parallelization is typically advocated to speed up the computing time, vectorization of the algorithm on a single processor may result in even larger speedups for certain problems. In this paper we present a formulation of the PF-SMC class of algorithms proposed in \cite{Arnold et al}, which is particularly amenable to a parallel or vectorized computing environment, and we illustrate the performance with a few computed examples in MATLAB.\\


\noindent \textbf{MSC-class}: 65Y05, 65Y10 (Primary); 62M20, 65L06, 62M05 (Secondary).
\end{abstract}

% Main content of the paper

\section{Introduction and motivation}
\label{sec:intro}

Parameter estimation for dynamic systems of nonlinear differential equations from noisy measurements of some components of the solution at discrete times is a common problem in many applications.  In the Bayesian statistical framework,  the particle filter (PF) is a popular sequential Monte Carlo (SMC) method for estimating the solution of the dynamical system and the parameters defining it in a sequential manner.  Among the different variants of PF proposed in the literature, the algorithm of \cite{Liu West} estimates the state variable along with the model parameters by combining an auxiliary particle technique \cite{Pitt Shephard 1999} with approximation of the posterior density of the parameter vector by Gaussian mixtures or an ensemble of particles drawn from the density \cite{West 1993a, West 1993b}.

Efficient time integration is crucial in the implementation of PF algorithms, particularly when the underlying dynamical system is stiff and cannot be solved analytically, therefore requiring the use of specialized numerical solvers.  In \cite{Arnold et al}, suitable linear multistep methods (LMMs) \cite{LeVeque, Iserles} for stiff problems are used within a \cite{Liu West}-type PF, and the variance of the innovation term in the PF is assigned according to estimates of the local error introduced by the numerical solver.  In the present work, we explain how to organize the calculations efficiently on multicore desktop computers, making it possible to follow a large number of particles.  Computed examples show that significant speedups can be obtained over a naive implementation.

The inherently parallel nature of PF algorithms is well-known (see, e.g., \cite{Brun et al, Maskell et al, Lee et al, Chopin et al}), and software packages have been made available for implementing parallel PFs on various platforms (e.g., \cite{Murray, Zhou}).  In this paper, we show how to reformulate the PF with LMM time integrators proposed in \cite{Arnold et al} to make it most amenable to parallel and vectorized environments.  The general approach can be straightforwardly adapted to different computing languages.  Computational advantages of the new formulations are illustrated with two sets of computed examples in MATLAB.

\section{An overview of the algorithm and a test problem}
\label{sec:pfsmc_review}

The derivation of the LMM PF-SMC method that we are interested in, inspired by the algorithm proposed in \cite{Liu West}, can be found in \cite{Arnold et al}.  For sake of completeness, the LMM PF-SMC procedure is outlined in Algorithm 1, where $\Psi$ denotes the LMM of choice.\\

\bigskip

{\footnotesize

\noindent \rule{\linewidth}{0.3mm}

\noindent {\bf Algorithm 1: LMM PF-SMC Sampler}

\noindent \rule{\linewidth}{0.3mm}

\noindent Given the initial probability density $\pi_0(x_0,\theta)$:

\begin{enumerate}
\item {\em Initialization:} Draw the particle ensemble from $\pi_0(x_0,\theta)$:
\begin{eqnarray*}
 &&{\mathcal S}_0 = \big\{(x^1_0,\theta_0^1,w_0^1),(x^2_0,\theta_0^2,w_0^2),\ldots,(x^N_0,\theta_0^N,w_0^N)\big\},\\
 && w_0^1 = w_0^2 = \ldots = w_0^N = \frac 1N.
\end{eqnarray*}
Compute the parameter mean and covariance:
\[
 \overline\theta_0 = \sum_{n=1}^N w_0^n\theta^n_0, \qquad
{\mathsf C}_0 =\sum_{n=1}^N w_0^n \big(\theta_0^n- \overline\theta_0\big)\big(\theta_0^n- \overline\theta_0\big)^{\mathsf T}.
\]
Set $j=0$.
\item {\em Propagation:} Shrink the parameters
\[
 \overline\theta_{j}^n = a\theta^n_j + (1-a)\overline\theta_j,\quad 1\leq n\leq N,
\]
by a factor $0<a<1$.
Compute the state predictor using LMM:
\[
 \overline x_{j+1}^n = \Psi(x_{j}^n,\overline\theta_{j}^n,h),\quad 1\leq n\leq N.
\]
\item {\em Survival of the fittest:} For each $n$:
\begin{itemize}
\item[(a)] Compute the fitness weights
\[
  g^n_{j+1} = w_j^n \pi(y_{j+1}\mid \overline x^n_{j+1},\overline\theta^n_{j}),\quad g^n_{j+1}\leftarrow \frac{g^n_{j+1}}{\sum_n g^n_{j+1}};
\]
\item[(b)] Draw indices with replacement
\[
\ell_n\in\big\{1,2,\ldots,N\big\}
\]
using probabilities ${\mathsf P}\{\ell_n = k\}=g^k_{j+1}$;
\item[(c)] Reshuffle
\[
  x_j^n \leftarrow x_j^{\ell_n}, \quad \big(\overline x^n_{j+1},\overline\theta^n_{j}\big) \leftarrow \big(\overline x^{\ell_n}_{j+1},\overline\theta_{j}^{\ell_n}\big), \quad 1\leq n\leq N.
\]
\end{itemize}
\item {\em Proliferation:} For each $n$:
\begin{itemize}
\item[(a)] Proliferate the parameter by drawing
\[
 \theta^n_{j+1}\sim{\mathcal N}\big(\overline\theta^n_{j},s^2{\mathsf C}_j\big),\quad s^2 = 1-a^2;
\]
\item[(b)] Using LMM error control, estimate
\[
{\mathsf \Gamma}^n_{j+1} = {\mathsf\Gamma}_{j+1}(x_j^n,\theta_{j+1}^n);
\]
\item[(c)] Draw $v^n_{j+1}\sim{\mathcal N}(0,{\mathsf\Gamma}_{j+1}^n)$;
\item[(d)] Repropagate using LMM and add innovation:
\[
  x_{j+1}^n = \Psi(x_{j}^n,\theta_{j+1}^n,h) + v^n_{j+1}.
\]
\end{itemize}
\item {\em Weight updating:} For each $n$, compute
\[
 w_{j+1}^n = \frac{\pi(y_{j+1}\mid x^n_{j+1},\theta^n_{j+1})}{\pi(y_{j+1}\mid \overline x^n_{j+1},\overline\theta^{n}_{j})},\quad w^n_{j+1}\leftarrow \frac{w^n_{j+1}}{\sum_n w^n_{j+1}}.
\]
\item If $j<T$, update
\[
\overline\theta_{j+1} = \sum_{n=1}^N w_{j+1}^n\theta^n_{j+1}, \quad
{\mathsf C}_{j+1} =\sum_{n=1}^N w_{j+1}^n \big(\theta_{j+1}^n- \overline\theta_{j+1}\big)\big(\theta_{j+1}^n- \overline\theta_{j+1}\big)^{\mathsf T},
\]
increase $j\leftarrow j+1$ and repeat from 2.
\end{enumerate}

\noindent \rule{\linewidth}{0.3mm}

}

\bigskip

The main computational bottlenecks in the implementation of the algorithm come from the numerical time integrations in Step 2 and Step 4, in particular when, due to the stiffness of the system, either extremely small time steps or the use of specially designed numerical schemes are needed to avoid the amplification of unstable modes.  Indeed, the need to use tiny time steps has been identified as a major bottleneck for PF algorithms; see, e.g., \cite{Calderhead et al 2009}.  Among the available ODE solvers for stiff systems, LMMs have the advantage being well-understood when it comes to stability properties and local truncation error estimates.  The latter, which in turn defines the accuracy of the integrator, and for which classical estimation methods exist, provides a natural way to assign the variance of the innovation term $v_{j+1}^n$ in Step 4; we refer to \cite{Arnold et al} for the details.

\subsection{A dynamical system with variable stiffness}

We illustrate how the organization of the computations affects the computing time of the proposed LMM PF-SMC algorithm on a system of nonlinear ODEs, which could arise, e.g., from multi-compartment cellular metabolism models \cite{Calvetti 2006, Arnold thesis, Golightly 2011}:
\begin{eqnarray}\label{diff eq system}
\Frac{dx_1}{dt} &=& \Phi(t) - V_1\Frac{x_1}{x_1+k_1} \nonumber \\
\noalign{\vskip4pt}
\Frac{dx_2}{dt} &=& V_1\Frac{x_1}{x_1+k_1} - V_2\Frac{x_2}{x_2+k_2}\\
\noalign{\vskip4pt}
\Frac{dx_3}{dt} &=& V_2\Frac{x_2}{x_2+k_2} - \lambda(x_3-c_0). \nonumber
\end{eqnarray}
The parameters $\lambda$ and $c_0$ are known, and
\[
\Phi(t) = A_0 + A(t-t_0)_+\exp(-(t-t_0)/\tau)
\]
is the input function, where $(t-t_0)_+$ is the non-negative part of $t-t_0$, and $A_0$, $A$, $t_0$, and $\tau$ are given.  In applications arising from metabolic studies, the components $x_1$, $x_2$ and $x_3$ of the solution of the ODE system, which will be referred to as the states of the system, are typically concentrations of substrates and intermediates.  In our computed examples, the data $y_j$ consist of noisy observations of all three state components $x_1$, $x_2$ and $x_3$ at 50 time instances, and the goal is to estimate the states at all time instances as well as the unknown parameters $V_j$ and $k_j$, $j=1,2$, which are, respectively, the maximum reaction rates and affinity constants in the Michaelis-Menten expressions of the reaction fluxes.  Since the system of ODEs \eref{diff eq system} is stiff for some values of the unknown parameters, we propagate and repropagate the particles using implicit LMMs, e.g., from the Adams-Moulton (AM) or backward differentiation formula (BDF) families.  Implicit methods require the solution of a nonlinear system of equations at each time step, which is done with a Newton-type scheme.  By carefully organizing the calculations so as to take maximal advantage of the multicore environment, it turns out that the time required by implicit LMM time integrators is comparable to that required by the explicit Adams-Bashforth (AB) integrators, which are not suitable for stiff problems.

\section{Parallel and vectorized formulations}
\label{sec:par or vec}

Many desktop computers and programming languages provide vectorized and multicore environments which can significantly reduce the execution time of PF methods when they are formulated to take advantage of these features.  All of the computed examples in this paper were produced using a Dell Alienware Aurora R4 desktop computer with 16 GB RAM and an Intel\textregistered\; Core\textsuperscript{\texttrademark} i7-3820 processor (CPU @ 3.60GHz) with 8 virtual cores, i.e., 4 cores and 8 threads with hyper-threading capability, using the MATLAB R2013a programming language.  When testing the parallel performance, we set the local cluster to have a maximum of 8 MATLAB workers, and we took as baseline the execution time of the LMM PF-SMC algorithm on a single processor.

\subsection{Parallel PF-SMC}
\label{subsec:parallelization}

It is straightforward to see that the propagation and re-propagation steps of Algorithm 1 are naturally suited to parallelization by subdividing the particles among the different processors.  This can be done by reorganizing the \verb"for" loops in the algorithm so that they are partitioned and distributed among the available processors (or workers) in the pool, which is achieved in MATLAB with the commands \verb"matlabpool" (or \verb"parpool") and \verb"parfor".

Not surprisingly, the best parallel performance occurs when all workers take approximately the same time to complete the task, because the slowest execution time determines the speed of the parallel loop.  This can be achieved by prescribing the same time step for all particles in the time integration procedure.  We remark that most ODE solvers, including the MATLAB built-in time integrators, guarantee a requested accuracy in the solution by adapting the time step, a practice which may cause the propagation of two different particles to take very different times, depending on the stiffness induced by different parameter values.  The spread of the computing times needed for the numerical integration of a particle ensemble is rather wide for systems, like the one in this example, whose stiffness is highly sensitive to the values of the unknown parameters.  This violates the principle of equal load on the workers which is essential for a good parallel performance.  Propagation of all particles by LMMs with the same fixed time step, on the other hand, ensures that the time required for each particle is the same, eliminating idle time.

To present the results of our computed examples, we introduce two key concepts in parallel computing: speedup and parallel efficiency.  The \emph{speedup} using $P$ processors is the ratio $S_P = T_1/T_P$, where $T_1$ is the execution time of the sequential algorithm and $T_P$ is the execution time of the parallel algorithm on $P$ processors, while the \emph{efficiency} using $P$ processors is defined as $E_P = S_P/P$.  Efficiency is a performance measure used to estimate how well the processors are utilized in running a parallel code: $E_P=1$, trivially, for algorithms run sequentially, i.e., on a single processor.  For further details, see, e.g., \cite{Scott et al}.

\begin{table}
\begin{center}
\begin{tabular}{|c|c c c c| c c c c|}
\hline
 & \multicolumn{4}{c}{$N=5,000$ particles} \hfill  \vline& \multicolumn{4}{c}{ $N=50,000$ particles}\hfill \vline \\
   \hline
LMM         & Sequential     & 8 Workers       & $S_8$       & $E_8$    & Sequential     & 8 Workers       & $S_8$       & $E_8$  \\
\hline
AB1         & 9.18e+02       & 4.83e+02        & 1.90        & 0.24     & 1.49e+04       & 3.95e+03        & 3.77        & 0.47     \\
AB2         & 1.67e+03       & 6.84e+02        & 2.44        & 0.31     & 2.24e+04       & 5.75e+03        & 3.90        & 0.49     \\
AB3         & 2.39e+03       & 8.72e+02        & 2.74        & 0.34     & 2.90e+04       & 7.48e+03        & 3.88        & 0.49   \\
\hline
AM1         & 5.15e+03       & 1.65e+03        & 3.12        & 0.39     & 5.63e+04       & 1.51e+04        & 3.73        & 0.47   \\
AM2         & 6.53e+03       & 2.05e+03        & 3.19        & 0.40     & 6.93e+04       & 1.93e+04        & 3.59        & 0.45    \\
AM3         & 7.79e+03       & 2.48e+03        & 3.14        & 0.39     & 8.26e+04       & 2.31e+04        & 3.58        & 0.45    \\
\hline
BDF1        & 4.14e+03       & 1.44e+03        & 2.88        & 0.36      & 4.63e+04       & 1.38e+04        & 3.36        & 0.42    \\
BDF2        & 4.74e+03       & 1.72e+03        & 2.76        & 0.35      & 5.35e+04       & 1.66e+04        & 3.22        & 0.40    \\
BDF3        & 5.47e+03       & 2.03e+03        & 2.69        & 0.34      & 5.84e+04       & 1.98e+04        & 2.95        & 0.37  \\
\hline
\end{tabular}
\end{center}

\caption{\label{par_table5}CPU times (in seconds) sequentially and in parallel with 8 workers, along with the corresponding speedup $S_8$ and efficiency $E_8$, for applying the PF-SMC algorithm to solve the parameter estimation problem for system \eref{diff eq system} using the first three LMM time integrators of each family with fixed time step $h = 0.05$ and two sample sizes, $N=5,000$ and $N=50,000$ particles, respectively.}
\end{table}

The results in Table \ref{par_table5} show that the use of parallel loops for test problem \eref{diff eq system} yields a significant speedup when using 8 workers, more pronounced for a sample of size $N=50,000$ than $N=5,000$ particles.  The efficiency is also higher for the larger sample size. Note that it takes longer to propagate with implicit (AM and BDF) solvers than with explicit (AB) solvers, both sequentially and in parallel, and that the CPU time increases with the order of the method.

We assign the number of workers in the above examples to match the number of virtual cores, which gives the best speedup for this problem on our local machine; using less workers (e.g., 4, the number of physical cores) may increase the parallel efficiency measure but yields less of a speedup in this case.

\subsection{Vectorized PF-SMC}
\label{subsec:vectorization}

Some of the most impressive reductions in computing time have been achieved by exploiting the capability of some higher level languages, including MATLAB, to perform certain operations in multithread mode, without the user having to explicitly open a pool of multiprocessors.  The number of built-in functions in MATLAB for which this holds continues to grow, including a wide range of operations between two- and three-dimensional arrays; therefore, algorithms which are formulated in a vectorized fashion are automatically candidates for internal parallelization.  It is important to note that in order to have a vectorized version of the PF algorithm, the time integration procedure must be identical for all particles.

In order to express the LMM PF-SMC algorithm for the system \eref{diff eq system} in a vectorized form for a sample of $N$ particles, we assemble a stacked column vector with $3N$ entries, corresponding to the three components of the state variables for the $N$ particles.  To maximize the number of vector-vector or matrix-vector operations, we implement the Newton-type method using an aggregate block diagonal matrix, whose diagonal blocks are the Jacobian matrices of the systems corresponding to each particle.  The resulting $3N \times 3N$ matrix contains mostly zeros, thus its sparse structure can also be exploited.  %Table \ref{vec_table5} lists the CPU times (in seconds) for solving the parameter estimation problem for system \eref{diff eq system} with vectorized PF-SMC using the first three LMM time integrators of each family with $N=5,000$ and $N=50,000$ particles, respectively.  The corresponding speedups over the sequential and parallel times listed in Table \ref{par_table5}, denoted by $S_\text{vec}^\text{seq}$ and $S_\text{vec}^{\text{par}8}$, are also reported.

\begin{table}
\begin{center}
\begin{tabular}{|c|c c c | c c c |}
\hline
 & \multicolumn{3}{c}{$N=5,000$ particles} \hfill  \vline& \multicolumn{3}{c}{ $N=50,000$ particles}\hfill \vline \\
 \hline
LMM     & Vectorized     & $S_\text{vec}^\text{seq}$    & $S_\text{vec}^{\text{par}8}$    & Vectorized     & $S_\text{vec}^\text{seq}$    & $S_\text{vec}^{\text{par}8}$ \\
\hline
AB1     & 6.51e+01       & 14.10           & \;7.42   & 6.05e+03       & \;2.46           & 0.65    \\
AB2     & 6.59e+01       & 25.34           & 10.38    & 6.00e+03       & \;3.73           & 0.96   \\
AB3     & 6.67e+01       & 35.83           & 13.07    & 6.01e+03       & \;4.83           & 1.24    \\
\hline
AM1     & 8.36e+01       & 61.60           & 19.74    & 6.19e+03       & \;9.10           & 2.44    \\
AM2     & 8.67e+01       & 75.32           & 23.64    & 6.19e+03       & 11.20            & 3.12    \\
AM3     & 8.64e+01       & 90.16           & 28.70    & 6.13e+03       & 13.47            & 3.77   \\
\hline
BDF1    & 8.35e+01       & 49.58           & 17.25    & 6.17e+03       & \;7.50           & 2.24   \\
BDF2    & 8.42e+01       & 56.29           & 20.43    & 6.14e+03       & \;8.71           & 2.70   \\
BDF3    & 8.53e+01       & 64.13           & 23.80    & 6.25e+03       & \;9.34           & 3.17    \\
\hline
\end{tabular}
\end{center}
\caption{\label{vec_table5}CPU times (in seconds) for solving the parameter estimation problem for system \eref{diff eq system} with vectorized PF-SMC using the first three LMM time integrators of each family with time step $h = 0.05$ and sample sizes of both $N=5,000$ and $N=50,000$ particles.  The corresponding speedups over the sequential and parallel times listed in Table \ref{par_table5}, denoted by $S_\text{vec}^\text{seq}$ and $S_\text{vec}^{\text{par}8}$, are also reported.}
\end{table}

The results in Table \ref{vec_table5} suggest that vectorizing the LMM PF-SMC algorithm for problem \eref{diff eq system} yields significant speedup over the sequential and even parallel implementations.  Moreover, the computing time for the vectorized LMM PF-SMC is essentially insensitive to the choice of LMM family, to the order of the method, and to the method used to estimate the discretization error.

\section{A dynamical system with uniform stiffness }

To test the performance of the parallelized and vectorized implementations of the PF-SMC algorithm on a different type of problem, we consider a two-dimensional (2D) advection-diffusion problem similar to that used in \cite{Lieberman Willcox 2013}, which can be thought of as modeling, e.g., the spreading of contaminants.  Consider the time-dependent partial differential equation
\begin{equation} \label{eq:advdiff}
\Frac{\partial u}{\partial t} = \nabla\cdot(\mathsf{D}\nabla u) + \mathbf{c}\cdot\nabla u
\end{equation}
over the square domain
\[
\Omega = [0,1]\times[0,1] = \{(x,y): 0\leq x\leq 1, 0\leq y \leq 1\}
\]
with periodic boundary conditions.  Here $\nabla u = \big(\frac{\partial u}{\partial x}, \frac{\partial u}{\partial y}\big)$, $\nabla\cdot$ is the divergence operator, $\mathsf{D}$ is a $2\times2$ matrix of constant diffusion coefficients and $\mathbf{c}$ is a $2\times1$ velocity vector describing the advection in the $x$ and $y$ directions.

The spatial domain is discretized into an $n\times n$ linearly spaced square grid.  At time $t=0$, $u$ is the sum of six Gaussian plumes,
\begin{equation}\label{eq:IC}
u_0(x,y) = \ds\sum_{i=1}^6 \Frac{1}{\gamma_i \sqrt{2\pi}}\exp\bigg\{-\Frac{1}{2\gamma_i^2}\big((x-x^c_i)^2 + (y-y^c_i)^2\big)\bigg\},
\end{equation}
whose standard deviations $\gamma_i$ and centers $(x^c_i,y^c_i)$ are listed in Table~\ref{Tab:IC}.

\begin{table}
\begin{center}
\begin{tabular}{|c|c c c c c c|}
\hline
$i$          & 1    & 2    & 3    & 4    & 5    & 6       \\
\hline
$\gamma_i$   & 0.04 & 0.08 & 0.07 & 0.10 & 0.05 & 0.06    \\
$x^c_i$      & 0.20 & 0.30 & 0.40 & 0.50 & 0.50 & 0.70    \\
$y^c_i$      & 0.80 & 0.40 & 0.40 & 0.50 & 0.60 & 0.50    \\
\hline
\end{tabular}
\end{center}
\caption{\label{Tab:IC} Standard deviations $\gamma_i$ and centers $(x^c_i,y^c_i)$ for the six Gaussian plumes  used to generate the simulated data for  \eref{eq:advdiff}.}
\end{table}

Since we assume periodic boundary conditions, we compute the solution on a grid of size $\widetilde{n} \times \widetilde{n}$, where $\widetilde{n} = n-1$, which we write as a stacked column vector $u$.
We assume that the $2\times2$ diffusion matrix
\[
\mathsf{D} = \left[ \begin{array}{cc} d_{11} & d_{12} \\ d_{21} & d_{22} \end{array} \right]
\]
is symmetric positive definite, and we write its Cholesky decomposition as $\mathsf{D} = \mathsf{K}^T\mathsf{K}$, where
\[
\mathsf{K} = \left[ \begin{array}{cc} k_1 & k_3 \\ 0 & k_2 \end{array} \right].
\]
It is straightforward to verify that
\[
d_{11} = k_1^2, \ d_{12} = d_{21} = k_1k_3, \mbox{ and } d_{22} = k_2^2 + k_3^2.
\]
We  want to estimate the parameters $k_1$, $k_2$, $k_3$, and the components $c_1$ and $c_2$ of the velocity vector $\mathbf{c}$.

To this end, we discretize the problem by introducing the matrices
\[
\Frac{\partial}{\partial x} \rightarrow B_1 = \mathsf{I}_{\widetilde{n}} \otimes \ell, \qquad \Frac{\partial}{\partial y} \rightarrow B_2 = \ell \otimes \mathsf{I}_{\widetilde{n}},
\]
where $\mathsf{I}_{\widetilde{n}}$ is the identity matrix of size $\widetilde{n} \times \widetilde{n}$, $\ell$ is an $\widetilde{n} \times \widetilde{n}$ finite difference matrix of the form
\[
\ell = \left[ \begin{array}{cccc} \ \ 1 & \ & \ & -1 \\ -1 & \ddots & \ & \ \\ \ & \ddots & \ddots & \ \\ \ & \ & -1 & \ \ 1 \end{array} \right]_{\widetilde{n} \times \widetilde{n}}
\]
and $\otimes$ denotes the Kronecker product, and obtain the discretized system of ODEs
\begin{equation} \label{eq:ODEsys}
\Frac{dU}{dt} = \mathsf{L} U,
\end{equation}
which we can propagate forward in time with LMM time integrators.  The operator matrix
\begin{eqnarray}\label{eq:operatormatrix}
\mathsf{L} &=& -\Big\{k_1^2B_1^T B_1 + k_1k_3(B_1^T B_2 + B_2^T B_1) \nonumber \\
 & & {} + (k_2^2 + k_3^2)B_2^T B_2\Big\} + c_1B_1 + c_2B_2
\end{eqnarray}
is given in terms of the five parameters of interest, where the matrices $B_1$, $B_2$, $B_1^TB_1$, $B_2^TB_2$ and $B_1^T B_2 + B_2^T B_1$ are sparse and need only be computed once; for details, see \cite{Arnold thesis}.  The inherent stiffness of system \eref{eq:ODEsys}, which requires the use of stiff solvers, comes from the spatial discretization of the original system and does not change significantly with the values of the parameters to be estimated; thus we generally do not expect the time for the propagation of the different particles to change as much as in the previous test problem.

\begin{figure}
\centerline{\includegraphics[width=1.5in]{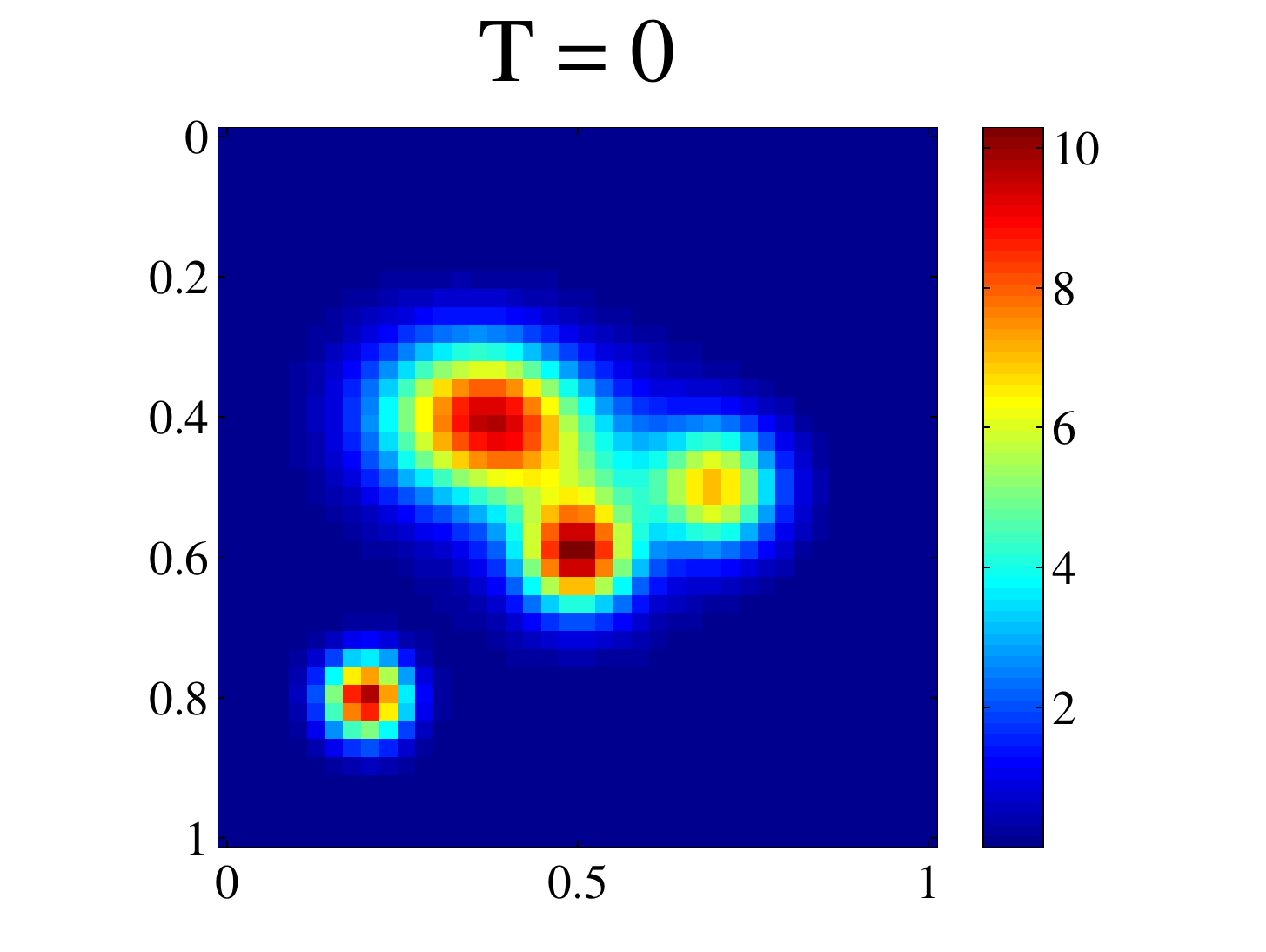} \includegraphics[width=1.5in]{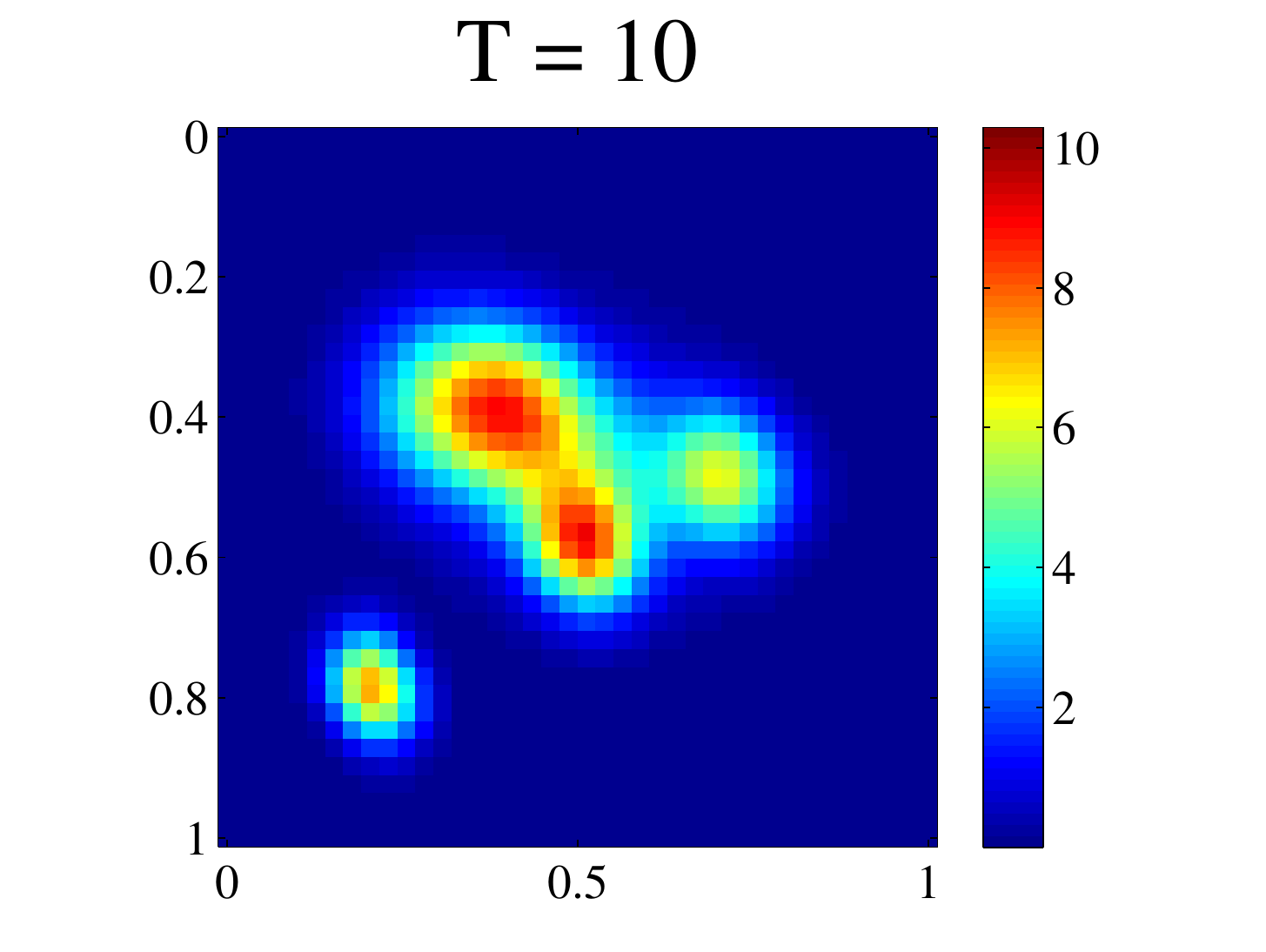} \includegraphics[width=1.5in]{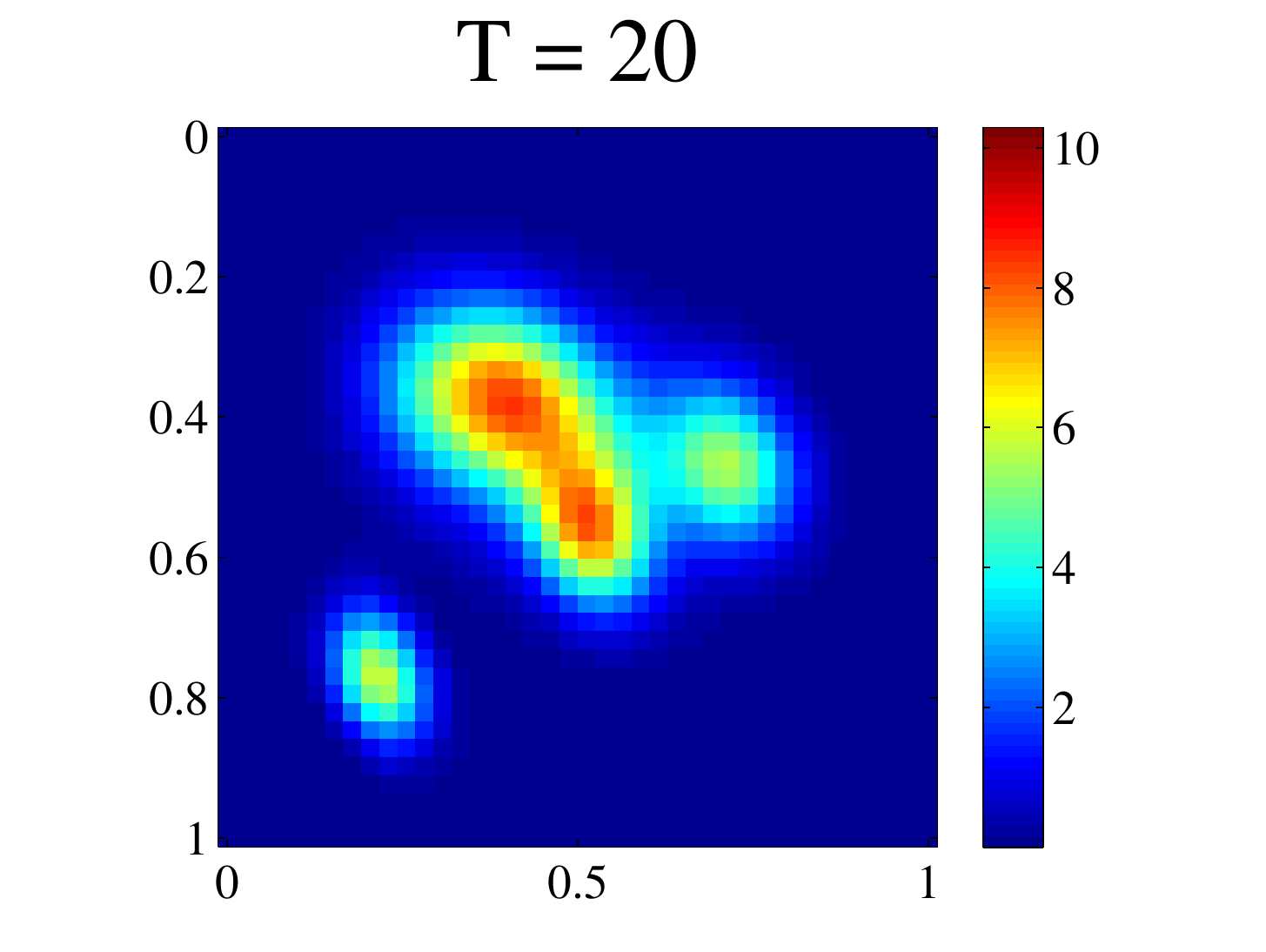} \includegraphics[width=1.5in]{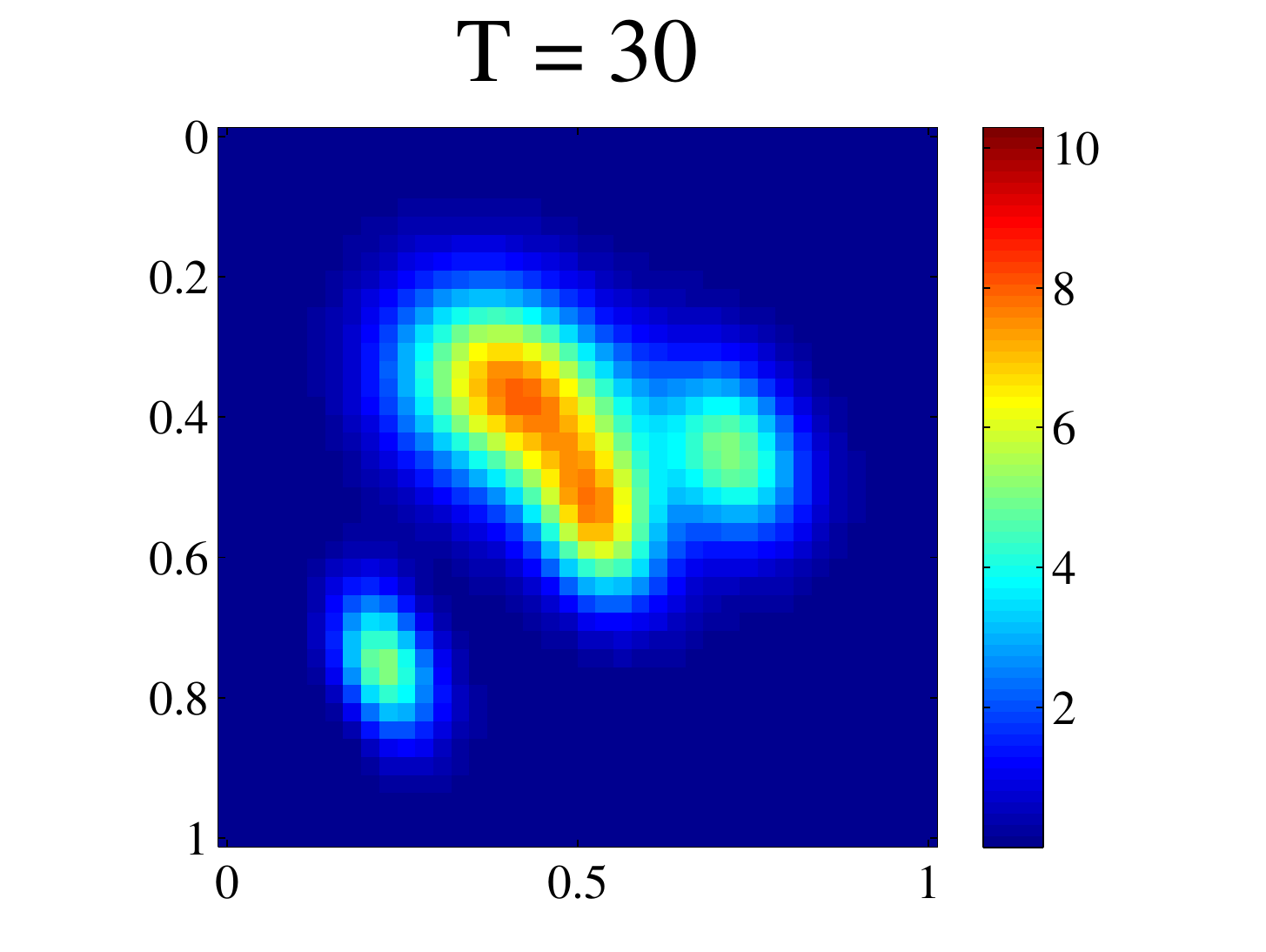}}
\caption{\label{Fig:advdiff sol} The initial image and reference solutions at 10, 20 and 30 seconds generated for problem \eqref{eq:advdiff} with discretization size $n=40$ and fixed parameter values given in \eref{fixedparams}.}
\end{figure}

We generate the data used by the PF by letting
\begin{equation}\label{fixedparams}
\mathsf{K} = \left[ \begin{array}{cc} 9 & 6 \\ 0 & 4 \end{array} \right] \quad \mbox{ and } \quad \mathbf{c} = \left[ \begin{array}{c} \ \ 2.5 \\ -1.5 \end{array} \right]
\end{equation}
and propagating \eref{eq:ODEsys} in MATLAB using the built-in stiff ODE solver \verb"ode15s" with relative tolerance $10^{-8}$ from $t=0$ to $t=30$ seconds, starting from the initial value computed according to \eref{eq:IC}.  The reference solution at 10, 20 and 30 seconds when $n=40$ is shown in Figure~\ref{Fig:advdiff sol}.  The data consist of the solution measured at 20 randomly selected, fixed spatial locations every one time unit, with Gaussian noise with standard deviation $\approx ||\text{data at time } t = 0||_{\infty}/50$ added to the observations, for a total of 30 noisy observations with 20 entries each.

\subsection{Parallelized vs. vectorized scheme}

Parallelization of the algorithm for system \eref{eq:ODEsys} is done along the lines of the example with variable stiffness in Section \ref{subsec:parallelization}.  Here the matrix $\mathsf{L}$ depends on the five parameters of interest and thus changes from particle to particle.  Since each particle can be propagated/repropagated independently, a separate operator matrix $\mathsf{L}$ defined by \eref{eq:operatormatrix} is constructed for each particle using its individual parameter values in the \verb"parfor" loop.

The vectorized implementation for this system requires considerable effort, since the $\mathsf{L}$ matrices for all of the $N$ particles, which are different because they depend on the parameters, must be constructed at once.  This can be achieved by building a sparse block diagonal matrix of size $N \widetilde{n}^2 \times N \widetilde{n}^2$ encompassing the operator matrices corresponding to all the of the individual particles, i.e.,
\begin{equation}\label{eq:vecL}
\mathsf{L} = \left[ \begin{array}{cccc} \mathsf{L}^{(1)} & \ & \ & \ \\ \ & \mathsf{L}^{(2)} & \ & \ \\ \ & \ & \ddots & \ \\ \ & \ & \ & \mathsf{L}^{(N)} \end{array} \right]_{N \widetilde{n}^2 \times N \widetilde{n}^2}
\end{equation}
where $\mathsf{L}^{(j)}$ is the $\widetilde{n}^2 \times \widetilde{n}^2$ operator matrix corresponding to the $j$th particle.

We remark that if we propagated the particles with an explicit AB method, the Jacobian computation would not be necessary, hence $\mathsf{L}$ would not need to be assembled, resulting in a sharp reduction in computing time.  However, AB methods are not well-suited for this class of stiff problems, as they becomes unstable unless a prohibitively small time step is used.

\begin{figure}
\centerline{\includegraphics[width=1.2in]{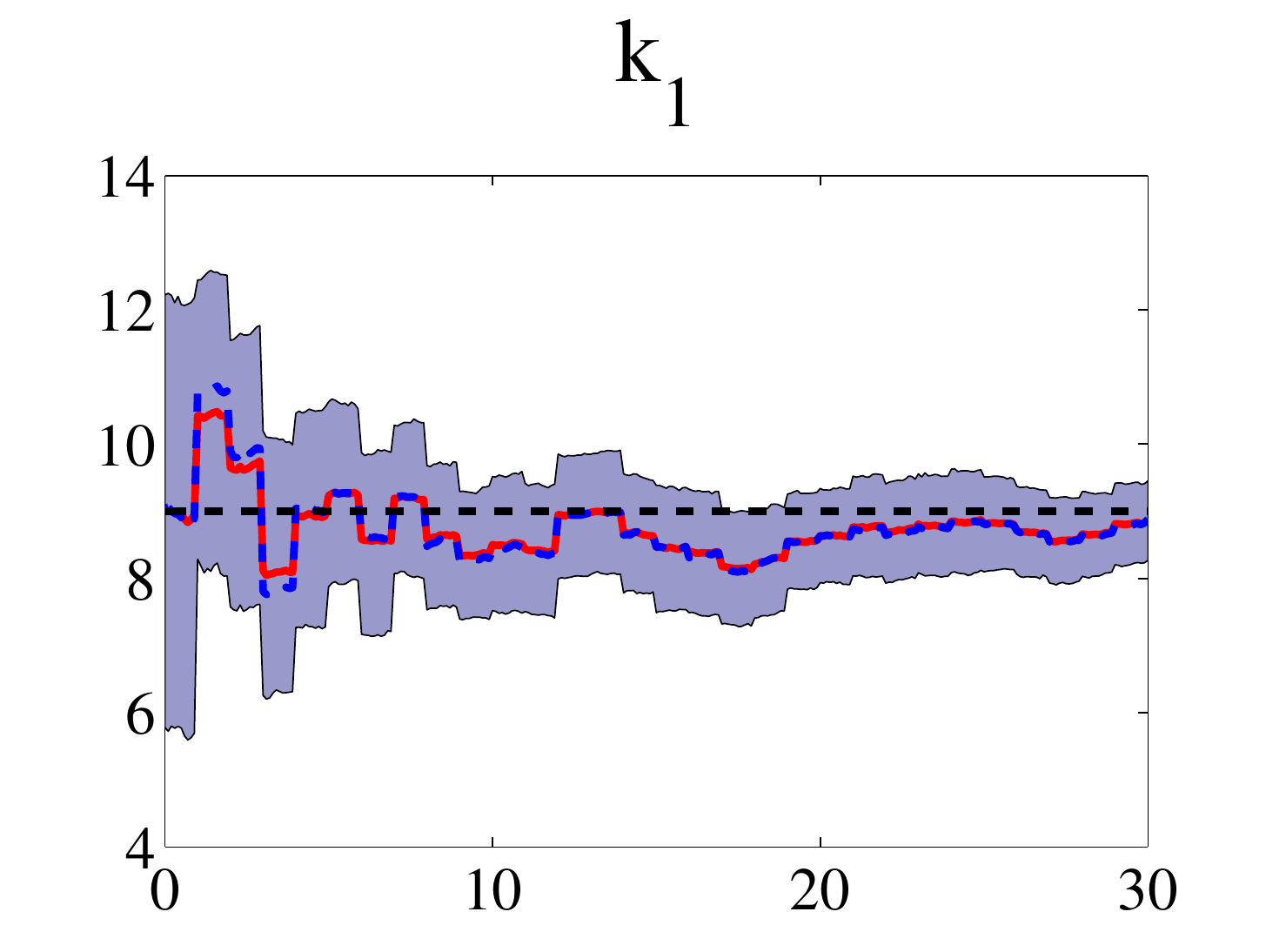} \includegraphics[width=1.2in]{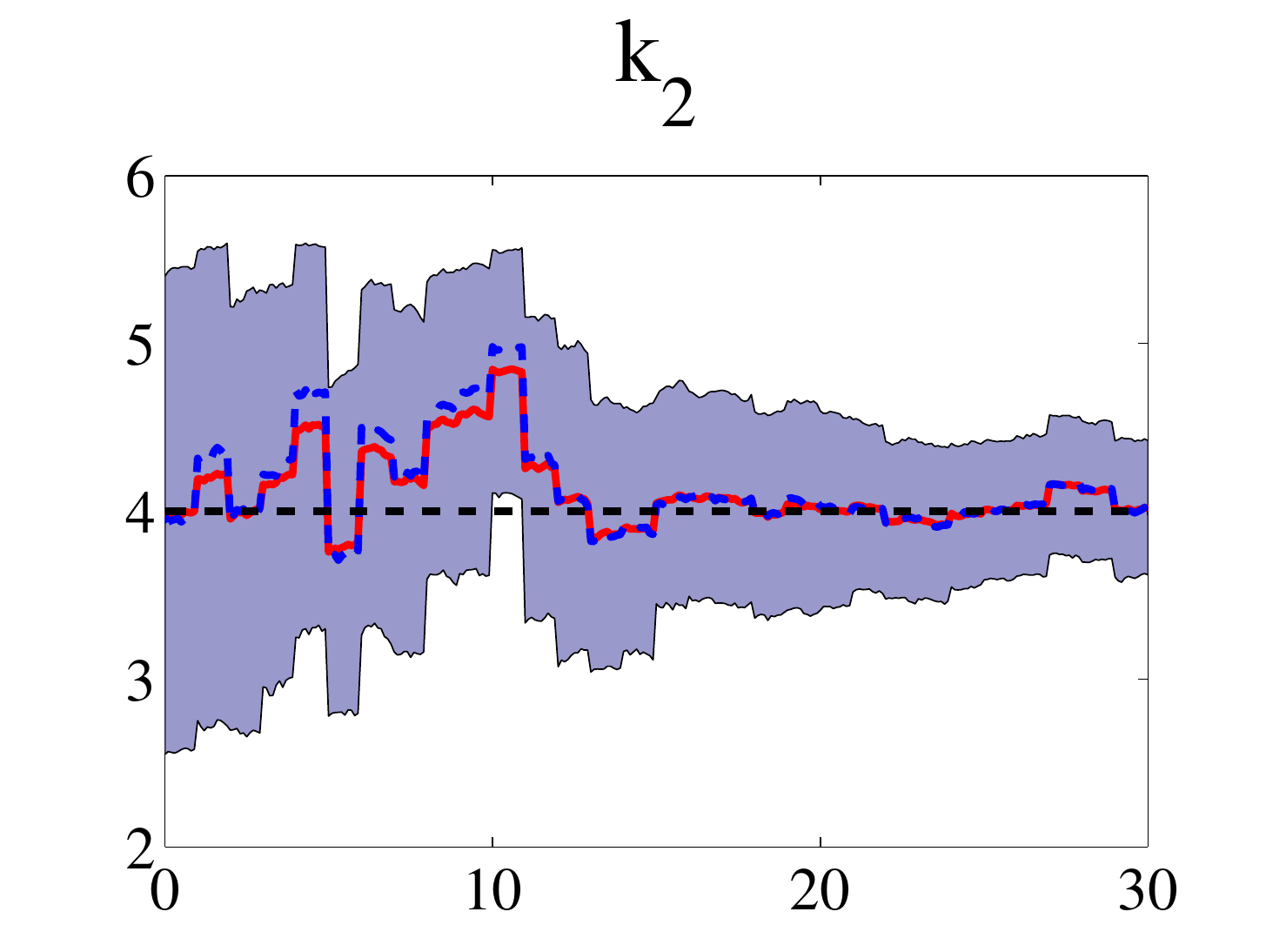} \includegraphics[width=1.2in]{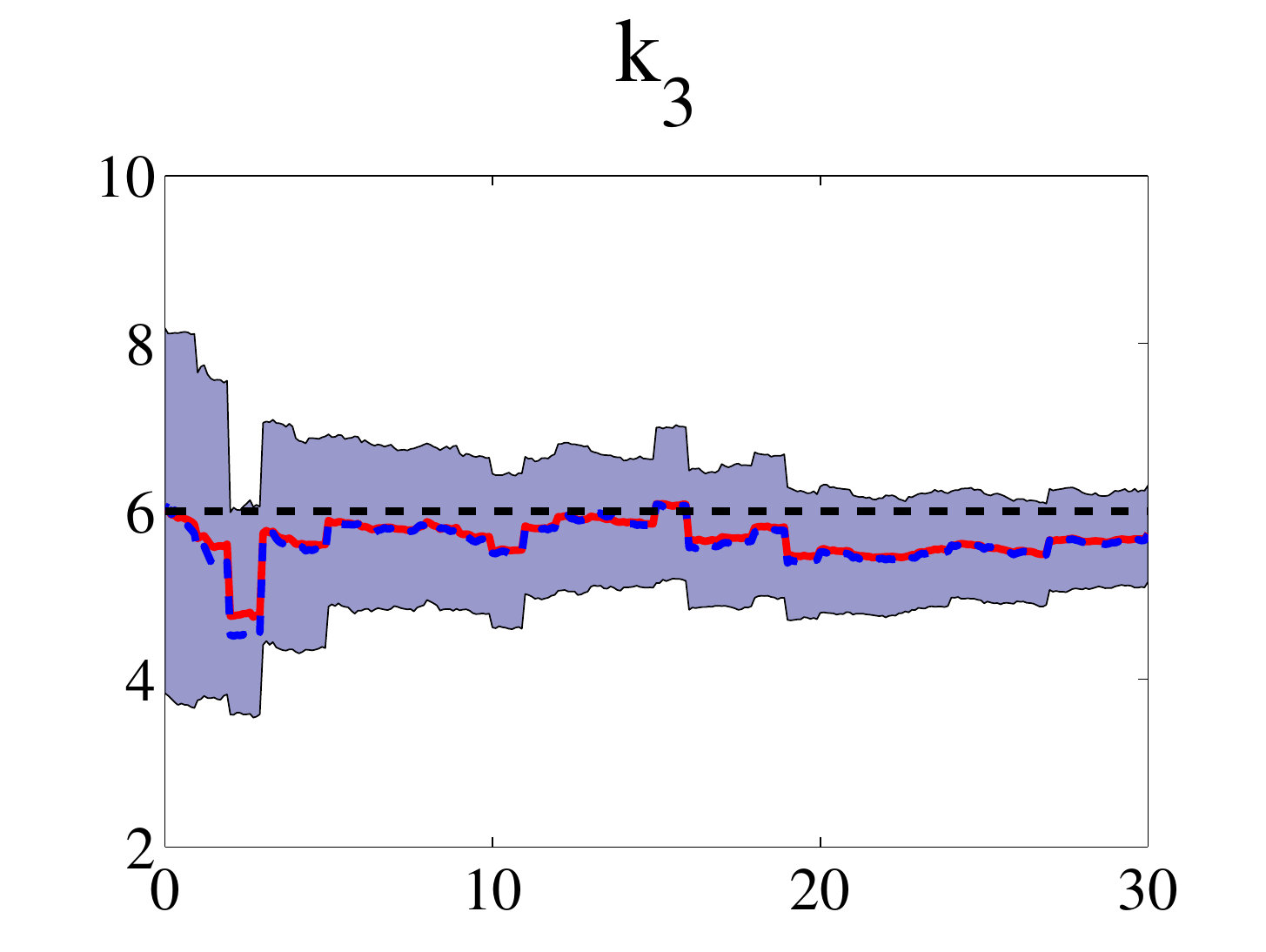} \includegraphics[width=1.2in]{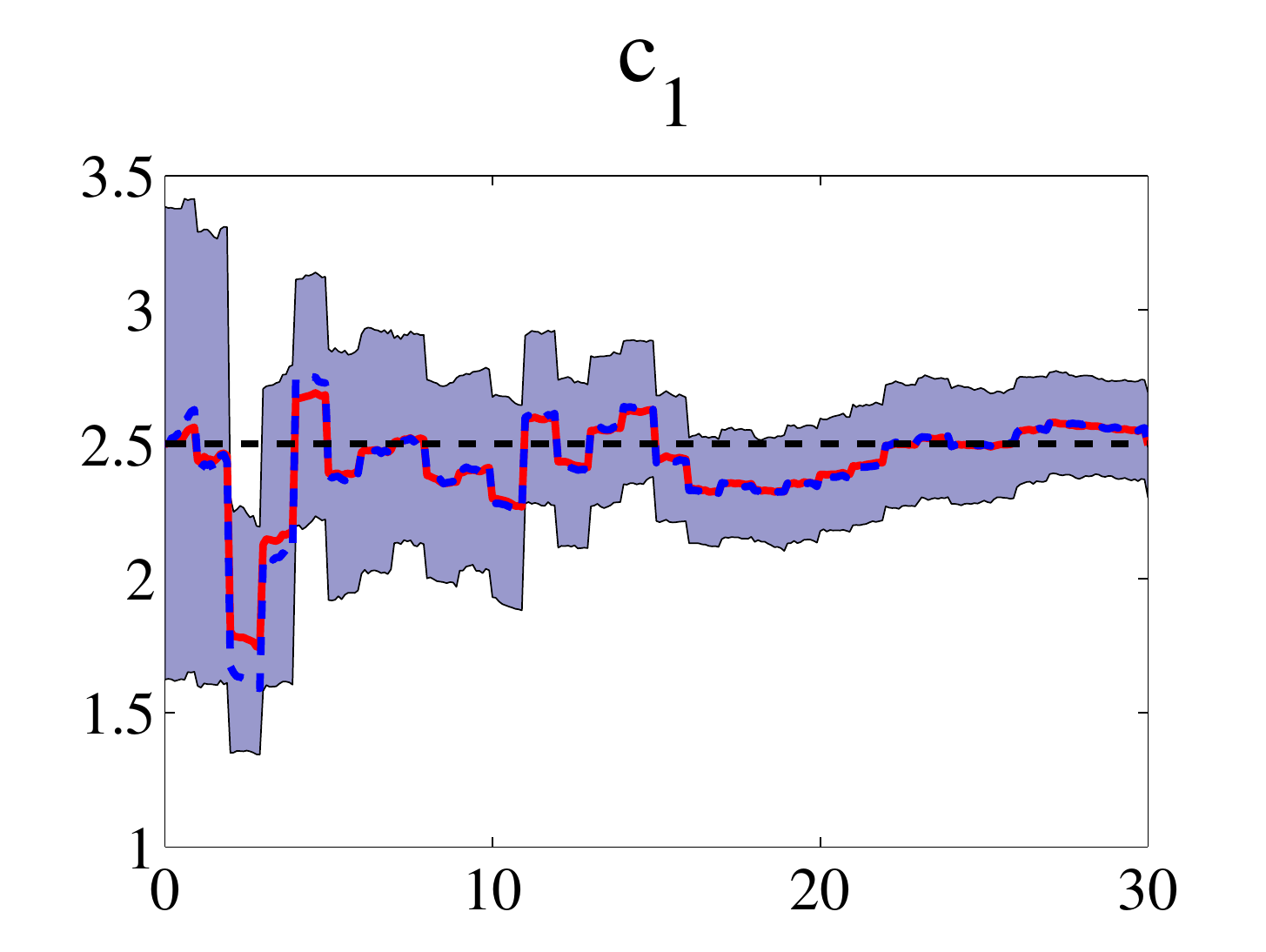} \includegraphics[width=1.2in]{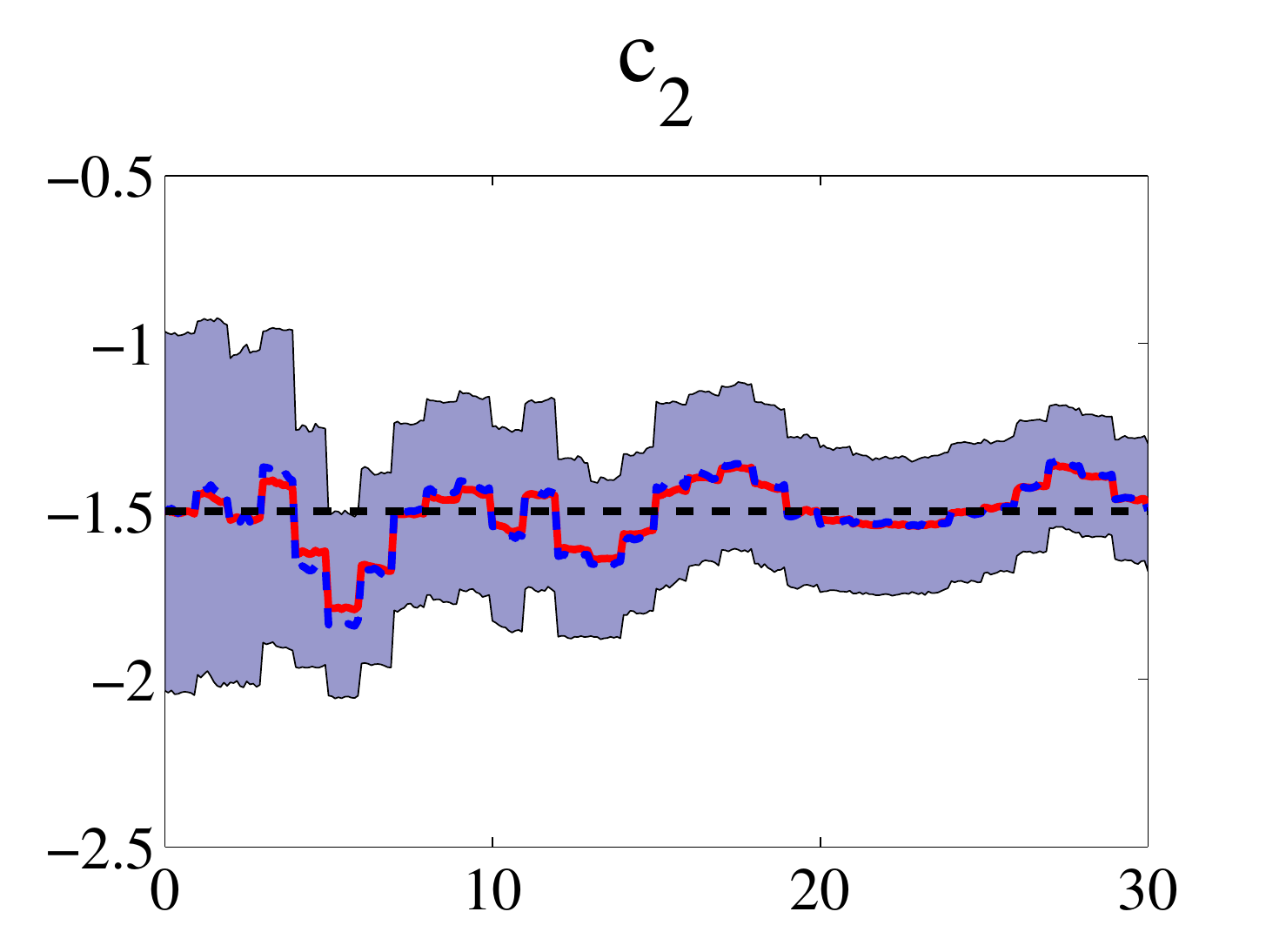}}
\caption{\label{Fig:advdiff results}Time series estimates of parameters $k_1$, $k_2$, $k_3$, $c_1$ and $c_2$ for problem \eref{eq:advdiff} when $n=20$.}
\end{figure}

The execution times for the sequential, parallel and vectorized implementations are listed in Table~\ref{Tab:advdiff}, where we also report the CPU times for sequential and parallel implementations of the PF-SMC where, instead of a fixed time step LMM, we use MATLAB's stiff solver \verb"ode15s" with the option to use BDF methods with maximum order 2 and default relative tolerance $10^{-3}$, assigning the innovation variance as the product of the relative tolerance and the number of integration steps taken by \verb"ode15s" from one datum arrival to the next.  Since \verb"ode15s" adjusts the time step as the integration proceeds, the number of steps taken may vary for each particle, but since the problem is uniformly stiff, it is not likely to vary too drastically.  Time series estimates of the parameters obtained when $n=20$ are shown in Figure~\ref{Fig:advdiff results}.

\begin{table}
\begin{center}
\begin{tabular}{|c|c c c|c| c c|}
\hline
 & \multicolumn{3}{c}{BDF2 with fixed time step} \hfill \vline & &\multicolumn{2}{c}{\texttt{ode15s} with fixed accuracy} \hfill \vline \\
 \hline
Size     & Sequential  & Parallel  & Vectorized  && Sequential  & Parallel  \\
\hline
$n=20$   & 3.08e+04    & 7.20e+03  & 3.26e+04   & & 1.26e+04    & 4.78e+03     \\
$n=40$   & 1.61e+05    & 3.39e+04  & 1.64e+05   & & 1.71e+05    & 8.77e+04      \\
\hline
\end{tabular}
\end{center}
\caption{\label{Tab:advdiff}CPU times (in seconds) when applying PF-SMC to system \eref{eq:ODEsys} of size $n$ sequentially, in parallel with 8 workers, and vectorized using BDF2 with time step $h = 0.1$ and $N=5,000$ particles (left) and for sequential and parallel implementations of PF-SMC using \texttt{ode15s} with fixed accuracy (right).}
\end{table}

\section{Discussion}

The use of stable, fixed time step LMM solvers in PF-SMC algorithms lends itself in a natural way to both parallelizing and vectorizing the computations, thus providing a competitive alternative to running independent parallel chains in Monte Carlo simulations \cite{Wilkinson 2005, Ren Orkoulas 2007}.  In this paper, we consider these two different implementation strategies for a recently proposed PF-SMC algorithm, and we illustrate the advantages with computed examples using two stiff test problems with different features.

The results in Tables \ref{par_table5} show that in the case where the stiffness of the dynamical system is very sensitive to the parameters to be estimated, as for system \eref{diff eq system}, vectorizing the LMM PF-SMC algorithm results in significant speedup over the sequential and even parallel implementations.  Moreover, for the vectorized version, the CPU times when using implicit and explicit methods are closer, and increasing the order of the method has little effect.

In the case of a large system where the stiffness is an intrinsic feature and does not depend much on the values of the unknown parameters, as for system \eref{eq:ODEsys}, on the other hand, vectorization of the PF-SMC using implicit LMMs does not perform better than the sequential implementation, but parallelization of the algorithm speeds up the calculations.  Our results suggest that both the size and structure of the problem determine whether the parallelized or vectorized version of the algorithm is more efficient.

% Acknowledgements section
\section*{Acknowledgments}
This work was partly supported by grant number 246665 from the Simons Foundation (Daniela Calvetti) and by NSF DMS project number 1312424 (Erkki Somersalo).

%%%%%%%%%%%%%%%%%%%%%%%%%%%%%%%%%%%%%%%%%%%%%%%%%%%%%%%%%%%%%%%%%%%%%%%%%%%%%%%%%%

% References

\end{document}